\documentclass[aps,onecolumn,preprint,superscriptaddress,nofootinbib,floats]{revtex4}
\usepackage{amsmath,amssymb,color,mathrsfs, graphicx,verbatim,epsfig, bbm, wasysym, axodraw}
\usepackage[hyperfootnotes=false]{hyperref}
\usepackage{slashed}
\usepackage{soul} 

\allowdisplaybreaks

\def\lsim{\mathrel{\rlap{\lower4pt\hbox{\hskip1pt$\sim$}}
    \raise1pt\hbox{$<$}}}
\def\gsim{\mathrel{\rlap{\lower4pt\hbox{\hskip1pt$\sim$}}
    \raise1pt\hbox{$>$}}}

\newcommand{\be}{\begin{eqnarray}}
\newcommand{\ee}{\end{eqnarray}}

\def\addresses#1#2{\hbox to \hsize{\@tablebox{#1}\hfil\@tablebox{#2}}}
\def\@tablebox#1{\vtop{\hsize=5in \begin{flushleft} #1 \end{flushleft}}}

\def\beq{\begin{equation}}
\def\eeq{\end{equation}}
\def\bit{\begin{itemize}}
\def\eit{\end{itemize}}
\def\beqa{\begin{eqnarray}}
\def\eeqa{\end{eqnarray}}

\def\MadGraph{{\tt MadGraph}}
\def\MadGraph5{{\tt MadGraph5}}

\newcommand{\met}{\displaystyle{\not}E_T}
\newcommand{\vecmet}{\vec{\displaystyle{\not}E}_T}
\newcommand{\mht}{\displaystyle{\not}H_T}
\newcommand{\vecmht}{\vec{\displaystyle{\not}H}_T}
\newcommand{\mpt}{\displaystyle{\not}p_T}

\begin{document}

\baselineskip 0.6cm

\begin{titlepage}

\thispagestyle{empty}

\begin{flushright}
PITT PACC 1508
\end{flushright}

\begin{center}

\vskip 1cm

{\Large \bf Revealing Compressed Stops Using \\ \vspace{2mm} High-Momentum Recoils}

\vskip 1.0cm
{\large Sebastian Macaluso$^{1}$, Michael Park$^{1,2}$, David Shih$^{1}$, and Brock Tweedie$^{3}$}
\vskip 0.4cm
$^{1}$ {\it NHETC, Department of Physics and Astronomy\\ Rutgers University, Piscataway, NJ 08854} \\
$^{2}$ {\it Stanford University, \\ Stanford, CA 94305} \\
$^{3}$ {\it PITT PACC, Department of Physics and Astronomy\\ University of Pittsburgh, Pittsburgh, PA 15260}
\vskip 2.0cm

\end{center}

\noindent  Searches for supersymmetric top quarks at the LHC have been making great progress in pushing sensitivity out to higher mass, but are famously plagued by gaps in coverage around lower-mass regions where the decay phase space is closing off. Within the common stop-NLSP / neutralino-LSP simplified model, the line in the mass plane where there is just enough phase space to produce an on-shell top quark remains almost completely unconstrained. Here, we show that is possible to define searches capable of probing a large patch of this difficult region, with S/B $\sim 1$ and significances often well beyond 5$\sigma$. The basic strategy is to leverage the large energy gain of LHC Run~2, leading to a sizable population of stop pair events recoiling against a hard jet. The recoil not only re-establishes a $\met$ signature, but also leads to a distinctive anti-correlation between the $\met$ and the recoil jet transverse vectors when the stops decay all-hadronically. Accounting for jet combinatorics, backgrounds, and imperfections in $\met$ measurements, we estimate that Run~2 will already start to close the gap in exclusion sensitivity with the first few 10s of fb$^{-1}$. By 300~fb$^{-1}$, exclusion sensitivity may extend from stop masses of 550~GeV on the high side down to below 200~GeV on the low side, approaching the ``stealth'' point at $m_{\tilde t} = m_{t}$ and potentially overlapping with limits from $t\bar t$ cross section and spin correlation measurements.

\end{titlepage}

\setcounter{page}{1}

\section{Introduction}
\label{sec:intro}

Light stops with mass below a TeV are extremely well-motivated by the supersymmetric solution to the hierarchy problem. The uniquely important role of these particles has inspired a growing and increasingly sophisticated set of dedicated searches at the LHC, targeting an array of different possible decay topologies~\cite{Aad:2014mfk,Aad:2014kra,Aad:2014nra,Aad:2014kva,Aad:2014bva,Aad:2014qaa,Aad:2014mha,Aad:2013ija,ATLAS_LRPVstop,CMScombinedStop,Khachatryan:2015wza,Khachatryan:2015pwa,Khachatryan:2014doa,Chatrchyan:2013xna,Chatrchyan:2013mya,CMSstopToCharm,Khachatryan:2014lpa} (see also~\cite{ATLASSUSY,CMSSUSY}). While these searches have already probed significant portions of the possible model space below a TeV, sizable gaps in coverage remain even at $O$(100~GeV), leaving us to consider: Is it possible that light stops have already been produced in abundance in LHC Run~1 but have simply been missed?

In perhaps the most minimalistic benchmark scenario, stops are produced directly in pairs via QCD, and each stop undergoes a one-step R-parity-conserving cascade into an invisible neutralino LSP and an on-shell or off-shell top quark:
\begin{equation}
pp\to \tilde t\tilde t^*,\qquad \tilde t\to t^{(*)}+\tilde\chi^0
\end{equation}
The visible composition of the final state is then identical to that of $t\bar t$, which serves as a copious background. The main kinematic handle exploited in most searches has been the additional injection of $\met$ (or more properly $\mpt$) from the neutralinos. For $m_{\tilde t} \gg m_{\tilde\chi}$, exclusion limits from $t\bar t + \met$ searches at Run~1 extend beyond 700~GeV~\cite{CMScombinedStop}. However, such searches face a major challenge when confronted with lower-mass regions in the stop-neutralino mass plane where the $\met$ is squeezed out. In particular, much attention has recently been directed at the ``top compression line'' $m_{\tilde t} \simeq m_{\tilde\chi} + m_t$, which defines the boundary between two-body decays into an on-shell top quark and neutralino, and three-body decays via an off-shell top quark into $Wb\tilde\chi^0$.  Limits along this compression line are largely nonexistent over a roughly 20~GeV-wide gap in stop mass.

Proposals to probe this region using the total $t\bar t$ cross section and spin correlations~\cite{Han:2012fw,Czakon:2014fka} have led to some inroads near the so-called ``stealth'' point $(m_{\tilde t},m_{\tilde\chi}) = (m_t,0)$~\cite{Aad:2014mfk,Aad:2014kva}, but theoretical limitations make it unclear if these searches can be pushed much further. The relatively long lifetimes of stops very near to the top compression line has led to a complementary suggestion to use the annihilation-decays of stoponium~\cite{Batell:2015zla,Kumar:2014bca,Kim:2014yaa}, which would lead to distinctive resonant diboson signatures (including, e.g., $\gamma\gamma$ and $Z\gamma$). Projections for Run~2 predict sensitivity up to stop masses of several hundred GeV, depending in detail on the stop chirality admixture. However, these searches become insensitive if the individual stops decay more quickly than the stoponium, which generally occurs as soon as the stop-neutralino mass difference opens up to even $O$(GeV). Other approaches have sought to use the small amount of $\met$ that is available within the bulk of the produced stop pair events. Very detailed measurements of the shapes of the tails of $\met$-sensitive observables may be promising~\cite{Alves:2012ft}, but a careful accounting of theoretical and experimental errors is not always available, and the one measurement of this type that has been carried out~\cite{Aad:2014kra} (by ATLAS, in the $l$+jets channel) does not reach the compression line. A simple cut-and-count style search based on dileptonic $m_{T2}$~\cite{Kilic:2012kw,Kats:2011qh} should still be viable due to a particularly sharp turnoff of the background, and is also sensitive near the stealth point. But the maximum mass reach of such a search is ultimately limited by low statistics and exhibits a significant dependence on stop chirality. It has also been suggested to utilize electroweak production of stop pairs via VBF, albeit with difficulties in probing stops much heavier than $m_t$~\cite{Dutta:2013gga,Buckley:2014fqa}.

Given these various limitations, there remains a clear need to consider further alternative options, lest comprehensive exploration of the top compression line be deferred to future precision lepton colliders. To make progress, we may take some inspiration from another compression line, at the very lowest end of the stop mass range: $m_{\tilde t} \simeq m_{\tilde\chi}$. There, not only the $\met$, but all visible activity is being squeezed out of the decay. Nonetheless, limits exist from the LHC, presently up to roughly 260~GeV~\cite{Aad:2014nra,CMSstopToCharm}. These are obtained using the classic trick of cutting into the region of production phase space where a sparticle pair is produced in association with a visible hard recoil particle, in this case a jet. For an almost completely compressed spectrum, the neutralinos go to zero velocity in the rest frames of their parents, but carry the full energy and therefore take up the full four-momenta. For stop pair production, the $\met$ vector in lab-frame is then automatically equal to the net $\tilde t \tilde t^*$ transverse momentum vector, which in turn approximately balances against the leading jet.

For stop-neutralino spectra near the top compression line, we can define an analogous trick, but now face several novelties. The neutralinos again approach zero velocity in their parent frames, but they share the four-momenta with (almost) on-shell sister top quarks, with fraction $m_{\tilde\chi}/m_{\tilde t}$ taken up by the neutralinos. Therefore, in the limit of perfectly compressed two-body decay $\tilde t \to t \tilde\chi^0$, and assuming a single dominant recoil jet with $\vec{p}_T({\rm jet}) \simeq -\vec{p}_T(\tilde t \tilde t^*)$, we get the following relation,
\beq
\vecmet \,\simeq\, -\vec{p}_T({\rm jet}) \times \frac{m_{\tilde\chi}}{m_{\tilde t}}.  \label{eq:recoilMET}
\eeq
The $\met$ is now attenuated relative to the recoil $p_T$, by a factor that can nominally extend down to zero in the massless neutralino limit (corresponding to the stealth stops~\cite{Han:2012fw}). This attenuation will generally make searches much more challenging when $m_{\tilde\chi} \ll m_{\tilde t}$ along the compression line, such that great care will be required in understanding the lower mass reach. For a given neutralino mass, the extra 2$m_t$ worth of energy required to make a stop pair also leads to much lower rates relative to conventionally compressed spectra with $m_{\tilde t} \simeq m_{\tilde\chi}$, especially in association with a proportionately energetic recoil jet. This issue in particular will be greatly ameliorated with the higher beam energy of the upgraded LHC. Finally, the two stop decays produce two on-shell or off-shell top quarks, which add to the visible activity and can inject further $\met$ if either $W$ decays leptonically. Perhaps somewhat counterintuitively, the cleanest signal may then be the all-hadronic decay mode, where all of the $\met$ comes from the neutralinos, and Eq.~\ref{eq:recoilMET} is most closely followed. However, this decay mode also maximizes possible QCD backgrounds, as well as our possible confusion over exactly which jets come from the recoil against the stop pair versus from their decays.

The possible utility of high-momentum recoils in this respect was emphasized relatively recently in~\cite{Hagiwara:2013tva}. In the present paper, we seek to put these ideas on firmer phenomenological footing, including a novel set of cuts and treatment of jet combinatorics, a detailed accounting of the various backgrounds, and allowance for a range of possible $\met$ measurement performances. Targeting all-hadronic stop decays, we typically find a healthy $S/B \sim 1$, ensuring robustness against systematic errors of up to $O(10\%)$. We proceed to make a detailed forecast for the possible discovery and exclusion coverage in the stop-neutralino mass plane. Our results are summarized in Fig.~\ref{fig:sensitivity}, where the proposed search is seen to cover a large portion of the formerly inaccessible top compression line, acting as a bridge between the two-body and three-body search strategies. For the expected 300~fb$^{-1}$ to be delivered through Run~3 of the LHC, exclusion sensitivity extends up to 550~GeV. On the lower end, shrinking $\met$ poses a major complication, but we find that exclusion sensitivity down to $m_{\tilde t} \simeq m_t + O(10$~GeV) may be possible. This would merge our forecasted coverage with that of $t\bar t$ cross section measurements and other techniques that perform well in the stealth region, allowing for unbroken coverage. If this can be achieved, it would be a major accomplishment of the LHC, and a further demonstration that the enormous luminosity and broad bandwidth of accessible energies there provides unique opportunities, even for relatively low-mass physics with subtle kinematics.

Our paper is organized as follows. The next section outlines our proposed analysis strategy and presents our estimated signal sensitivities. Section~\ref{sec:conclusions} discusses the results and possible extensions. More details of the generation of our event samples are presented in Appendix~\ref{sec:generation}.

\section{Proposed Analysis and Predicted Coverage}
\label{sec:analysis}

Our proposed analysis requires only a few ingredients:
\begin{itemize}
\item  A veto on isolated leptons.
\item  A high multiplicity of jets and at least two $b$-tags.
\item  An energetic ``ISR-jet'' candidate. 
\item  Coarsely-reconstructed top-candidates whose masses are not significantly above $m_t$.
\item  A strong anticorrelation of ISR-jet and $\vecmet$ directions.
\item  A ``significant'' amount of missing energy, $\met/\sqrt{H_T}$, localized near a value set by the ISR-jet $p_T$ cut and $m_{\tilde\chi}/m_{\tilde t}$.
\end{itemize}
In more detail, our full reconstruction and selection, applied to 13~TeV simulated data (Appendix~\ref{sec:generation}), proceeds as follows.

Reconstructed electrons (muons) are first selected starting from truth leptons with $p_{T,\ell} > 10~$GeV and $|\eta_{\ell}| < 2.5$, and flat identification efficiency of $0.90$ ($0.95$). Electrons are then isolated by first computing $\sum_i | p_{T,i} |_{\Delta R < 0.2}$ (where the sum is over all other particles within $\Delta R < 0.2$ of the electron) and requiring
\beq
\frac{\sum_i | p_{T,i} |_{\Delta R < 0.2}}{p_{T,\ell}} < 0.1 \, .
\eeq
Electrons that fail this isolation criterion, as well as all other unidentifiable leptons, are returned to the particle list as ``hadrons'' to be used in jet clustering. Additionally, there must be no jets (defined below) within $0.4$ of either an electron or muon. Otherwise, the lepton is vector-summed into the closest jet.\footnote{While these steps do not explicitly fold in pileup, significant drops in lepton reconstruction and isolation efficiencies in the coming LHC runs are unlikely, especially given the availability of isolation methods that are more tracker-based. It is also important to note that, because of the high recoil $p_T$ cut demanded below, leptons in the dominant backgrounds tend to be quite energetic.} Events that contain any surviving isolated leptons are then discarded. This lepton veto significantly reduces important backgrounds where the $\met$ arises from a $W$ boson decay, especially $l$+jets $t\bar t$ events and leptonic $W$+jets. More aggressive approaches than ours are also possible, using $\tau$ anti-tagging and/or vetoes on more loosely-identified leptons. Ultimately, we find that our backgrounds containing $W$s are moderately dominated by $\tau\nu_\tau$.

Jets are clustered from all truth hadrons, photons, and unidentified leptons (including electrons that fail the initial isolation step). The anti-$k_T$ algorithm~\cite{Cacciari:2008gp} in {\tt FastJet}~\cite{Cacciari:2005hq} is applied with $R = 0.4$, an initial $p_T$ threshold of 15~GeV, and $|\eta| < 5.0$. Jets from this stage are used for the lepton isolation above. Individual jet energies are then smeared with gaussians according to the expectation for the Run~2 \&~3 conditions of $\approx 50$ simultaneous pileup events, as projected in the Snowmass 2013 simulation note~\cite{Anderson:2013kxz}: $\sigma(p_T)/p_T = (8.2~{\rm GeV})/p_T \oplus (0.55~{\rm GeV}^{1/2})/\sqrt{p_T} \oplus 0.02$.\footnote{As of this writing, the most recent version (v1) contains a shifted-decimal typo for the noise coefficient in the written formula.} Subsequently, an event must have at least seven reconstructed jets with smeared $p_T$ above 20~GeV, highly favoring the all-hadronic $\tilde t \tilde t^*$+jet signal topology and further reducing backgrounds.\footnote{We do not model ``pileup jets'' consisting mostly of diffuse pileup particles, of which $O(2)$ per event are expected~\cite{ATLAS_HL-LHC} given our $p_T$ threshold and before dedicated pileup-jet rejection. We anticipate that these will be rejected with reasonable enough efficiency (see, e.g.,~\cite{ATLAS_pileup}) so as not to have a major impact on our analysis, though higher thresholds on the individual jet $p_T$s would also be an option if necessary.}

Jets with $|\eta| < 2.5$ are $b$-tagged according to an assumed working point with an efficiency of $0.70$ ($0.10$) for truth $b$-jets ($c$-jets). Jets are first truth flavor-tagged by looking for the heaviest overlapping $b$- or $c$-hadron in the event record, and then assigned a reconstruction-level identity ($b$-jet or light-flavor jet) based on the above efficiencies. Mistags of light-flavor jets are not incorporated, nor are backgrounds with less than two heavy-flavor partons in the hard event (see Appendix~\ref{sec:generation}). Light-flavor mistags are of subleading importance for both the stop signal and top backgrounds. For $W/Z$+jets and especially multijets, a complete analysis with light-flavor mistags requires extensive simulation, which we have not undertaken. However, we do not expect this omission to have significant impact on the validity of our background estimates. As a specific corroborating example, we refer to the detailed background composition of the Higgs search $(W/Z)H\to (W/Z)(b\bar b)$~\cite{ATLAS_detailedVH}, in which the $W/Z$+jets backgrounds are dominated by events with two truth $b$-jets.\footnote{To give some rough sense of accounting, the ``penalty'' for QCD to produce a pair of hard, well-separated heavy quarks from a gluon splitting is $O(\alpha_s/\pi)$, which is overall percent-scale. This easily beats the chances of a double-mistag of truth light-flavor jets, which is $O(10^{-4})$. For single-mistag events containing one $b$-quark at the hard event level, the $O(10^{-2})$ mistag would need to be combined with the very small $b$ PDFs. (Practically such events are paying both $\alpha_s/\pi$ {\it and} the mistag rate.)}

The $\met$ vector is modeled in two ways, with the hopes of bracketing true performance. The first model is truth $\vecmet$, which though simple to define in simulation is clearly overly optimistic. The second model is $\vecmht \equiv -\sum_j \vec{p}_T(j)$. This definition is technically somewhat pessimistic because it does not account for corrections that may come from activity not clustered into jets. However, we re-emphasize that pileup effects have been incorporated into the jet energy resolutions. For both definitions, $\vecmet$ ($\vecmht$) is not allowed to point along the $\vec p_T$ of any of the leading three jets, with a requirement $|\Delta\phi| > 0.55$. In practice, such a cut is used experimentally to avoid fake $\met$ from under-measured jets, as well as real $\met$ from heavy flavor decays inside of jets. Within our own multijets samples, the cut is still somewhat advantageous when using $\mht$. The advantage with truth $\met$ is minor, but we continue to apply the cut to maintain consistency and a higher degree of realism.

\begin{table}[t!]
\centering
\begin{tabular}{|l|c|}
\hline
\ lepton veto          \ & \ no isolated ID'ed leptons with $p_T(l) > 10$~GeV, $|\eta(l)| < 2.5$ \ \\ \hline
\ jets                 \ & \ $p_T(j) > 20$~GeV, $|\eta(j)| < 5$; $N(j) >= 7$, $N(b$-tag$) >= 2$ \ \\ \hline
\ ISR-jet              \ & \ $p_T$(ISR-jet)$\, > 550$~GeV \ \\ \hline
\ tops                 \ & \ $m$(top-candidates)$\, < 250$~GeV \ \\ \hline
\ jet/$\met$ alignment \ & \ $|\Delta\phi(j_{1,2,3},\met)| > 0.55$, $|\Delta\phi($ISR-jet,$\met)| > 2.95$ \ \\ \hline
\ $\met/\sqrt{H_T}$    \ & \ optimized window (minimally $\mht/\sqrt{H_T} > 3$~GeV$^{1/2}$) \ \\ \hline
\end{tabular}
\caption{Summary of reconstruction cuts.} 
\label{tab:cuts}
\end{table}

Identification of the ISR jet exploits the kinematics of top decay in a simple way. For a $b$-quark produced in a hadronic top decay, adding in either of the quarks produced in the sister $W$'s decay will produce a subsystem with a mass less than $m_t$, and more specifically less than $\sqrt{m_t^2-m_W^2} \simeq 153$~GeV at leading-order with narrow $W$. These inequalities continue to hold even when the top is below its mass-shell, as the kinematic boundary only becomes lower. The leading two $b$-jets in the event are taken to be the $b$-quark candidates. A list of remaining jets in the event is formed which satisfy $m(b+j) > 200$~GeV for {\it both} $b$-quark candidates. The highest-$p_T$ jet from this list is then the ISR candidate. Only events with $p_T$(ISR-jet)$\, > 550$~GeV are kept in our analysis.

\begin{figure*}[t!]
\begin{center}
\includegraphics[width=0.44\textwidth]{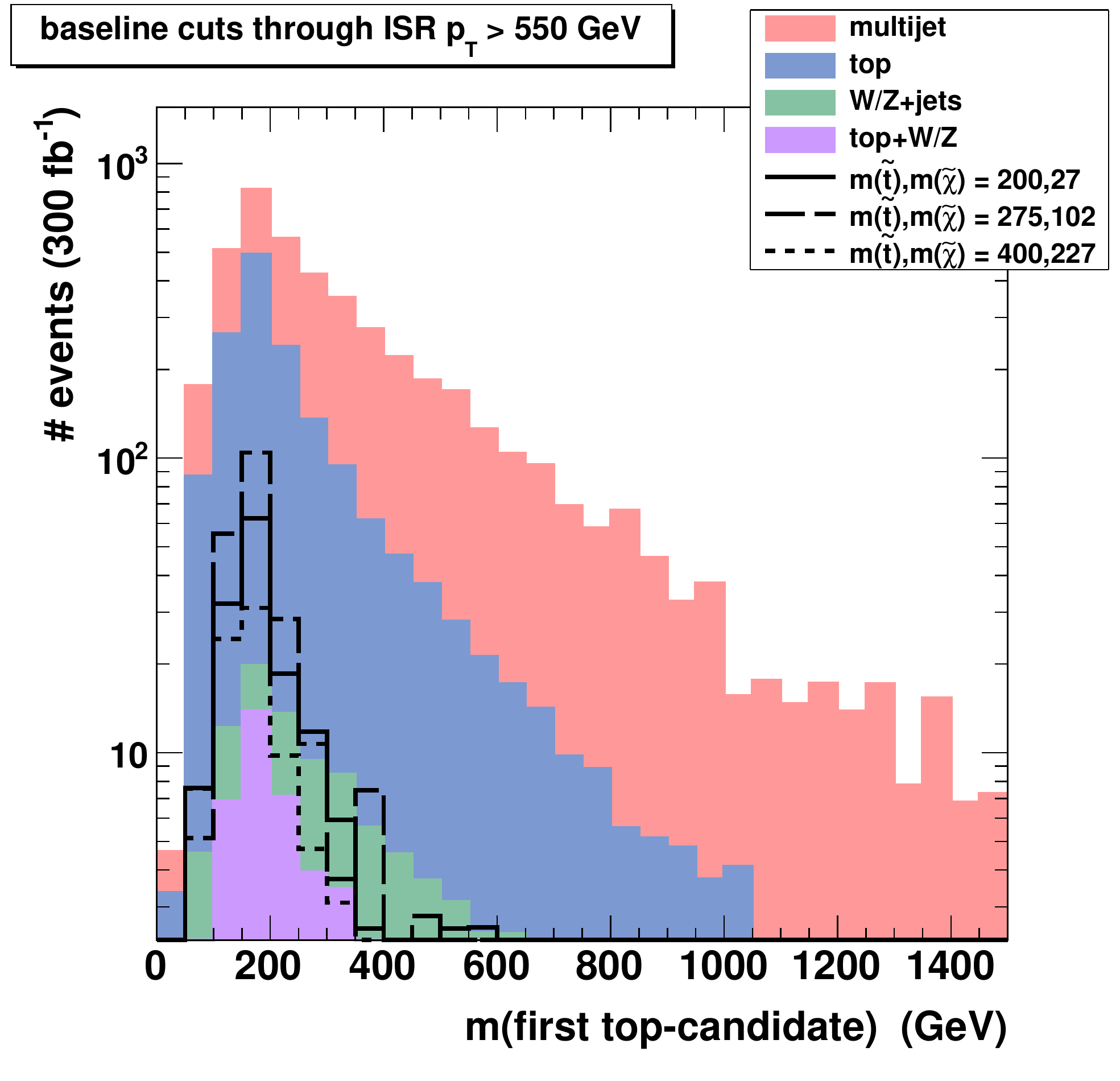} \hspace{0.2cm}
\includegraphics[width=0.44\textwidth]{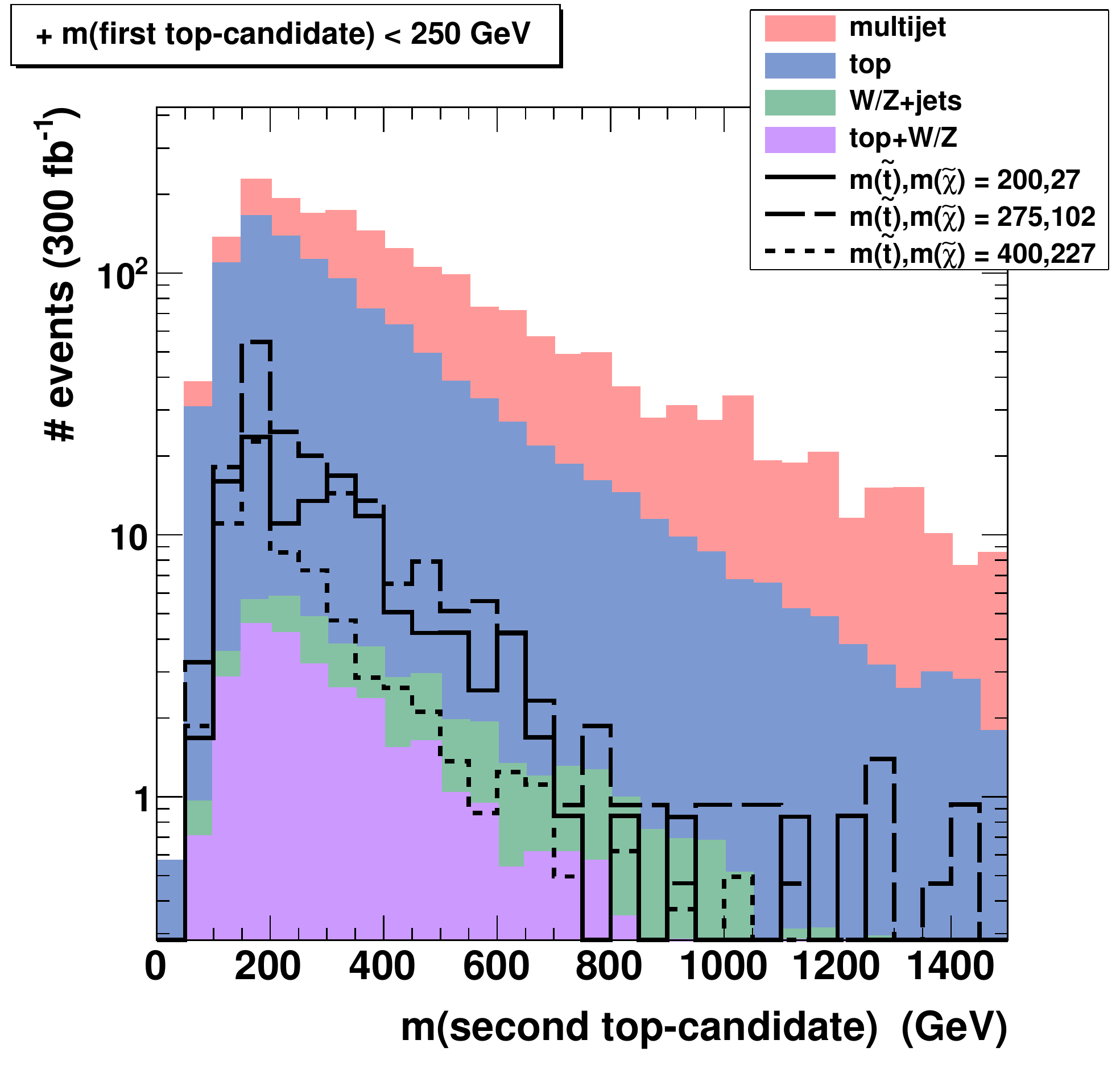}\\
\includegraphics[width=0.44\textwidth]{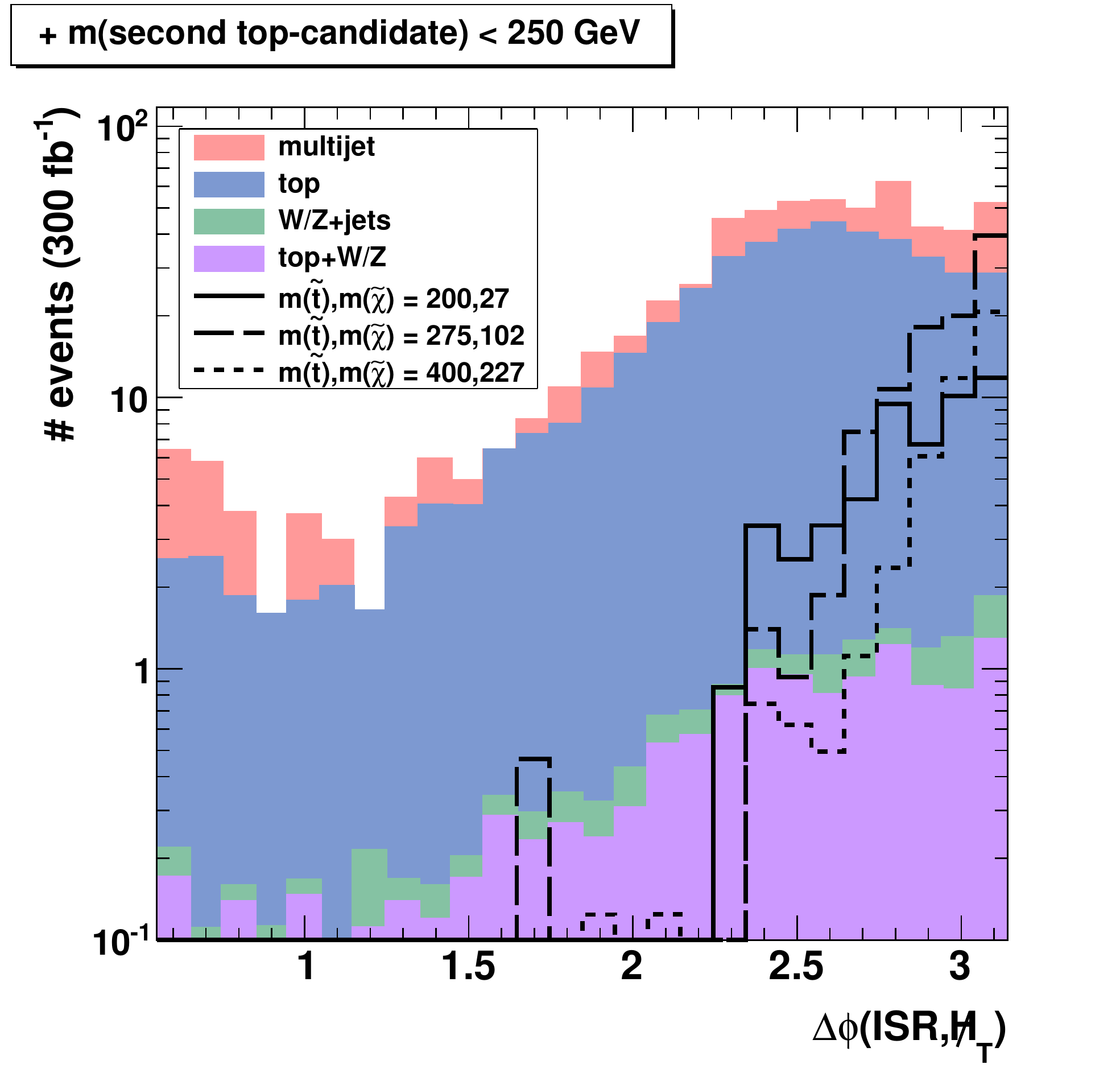}
\caption{Kinematic distributions of stacked backgrounds and some example signal points for the $\mht$-based analysis, with cumulative cuts. The baseline cuts include the lepton veto, jet counting, ISR-jet $p_T$ cut, and a cut $\mht/\sqrt{H_T} > 3$~GeV$^{1/2}$ used to define the simulation samples.}
\label{fig:mt1}
\end{center}
\end{figure*}

Individual top quarks are reconstructed using a procedure borrowed from~\cite{Aad:2014bva}. Excluding the two leading $b$-jets and the ISR-jet candidate, the two closest jets in the $\eta$-$\phi$ plane are added to form a ``$W$ boson.'' This in turn is added to the closest $b$-jet to form a ``top quark.'' The procedure is then repeated amongst the remaining jets and $b$-jet. In the absence of smearings and combinatoric confusions, both top-candidates constructed in this manner would satisfy $m \simeq m_t$ if on-shell, and $m < m_t$ if off-shell. We make a somewhat looser demand of $m < 250$~GeV. The main purpose of this cut is to reduce multijet and $W/Z$+jet backgrounds, which tend to reconstruct higher masses with a very broad tail.

Finally, we employ the relation in Eq.~\ref{eq:recoilMET}, which, as per~\cite{Hagiwara:2013tva}, we decompose into angle and magnitude. For the angular component, a strong anticorrelation between the ISR-jet and $\met$ directions is demanded: $|\Delta\phi($ISR-jet,$\met)| > 2.95$. For the magnitude, we expect that the signal $\met$ will be approximately equal to $p_T$(ISR-jet)$\times (m_{\tilde\chi}/m_{\tilde t})$. Because of the interplay of the hard $p_T$(ISR-jet) cut and the rapidly-falling production $p_T$ distributions, the signal will appear as a localized bump in $\met$.  Raw $\met$ can serve as an adequate discriminating variable here, but we instead use the ``significance'' ratio $\met/\sqrt{H_T}$, which we find to be slightly more efficient at separating signal from background for the lower-mass signals.

\begin{figure*}[tp!]
\begin{center}
\includegraphics[width=0.44\textwidth]{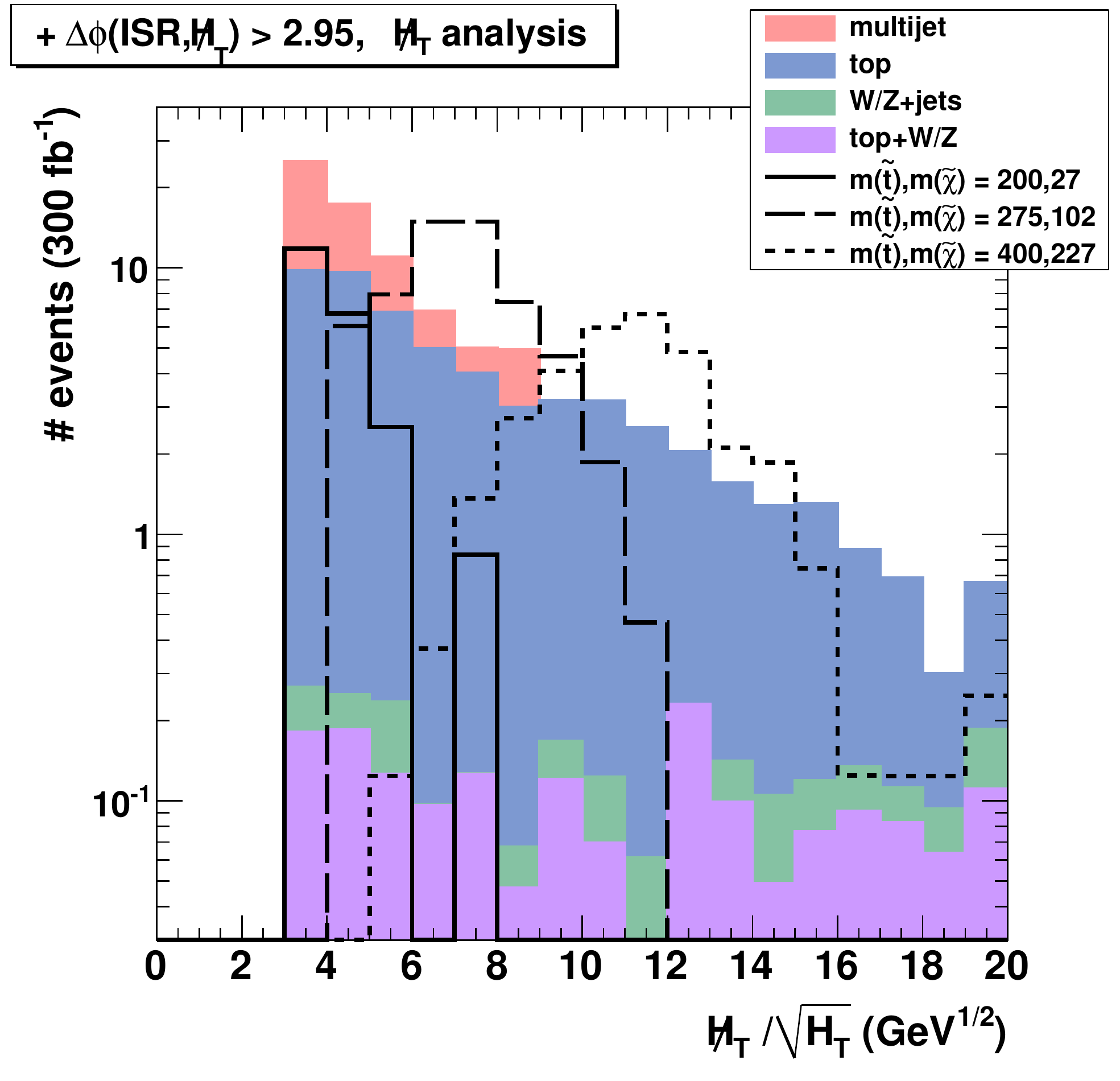} \hspace{0.2cm}
\includegraphics[width=0.44\textwidth]{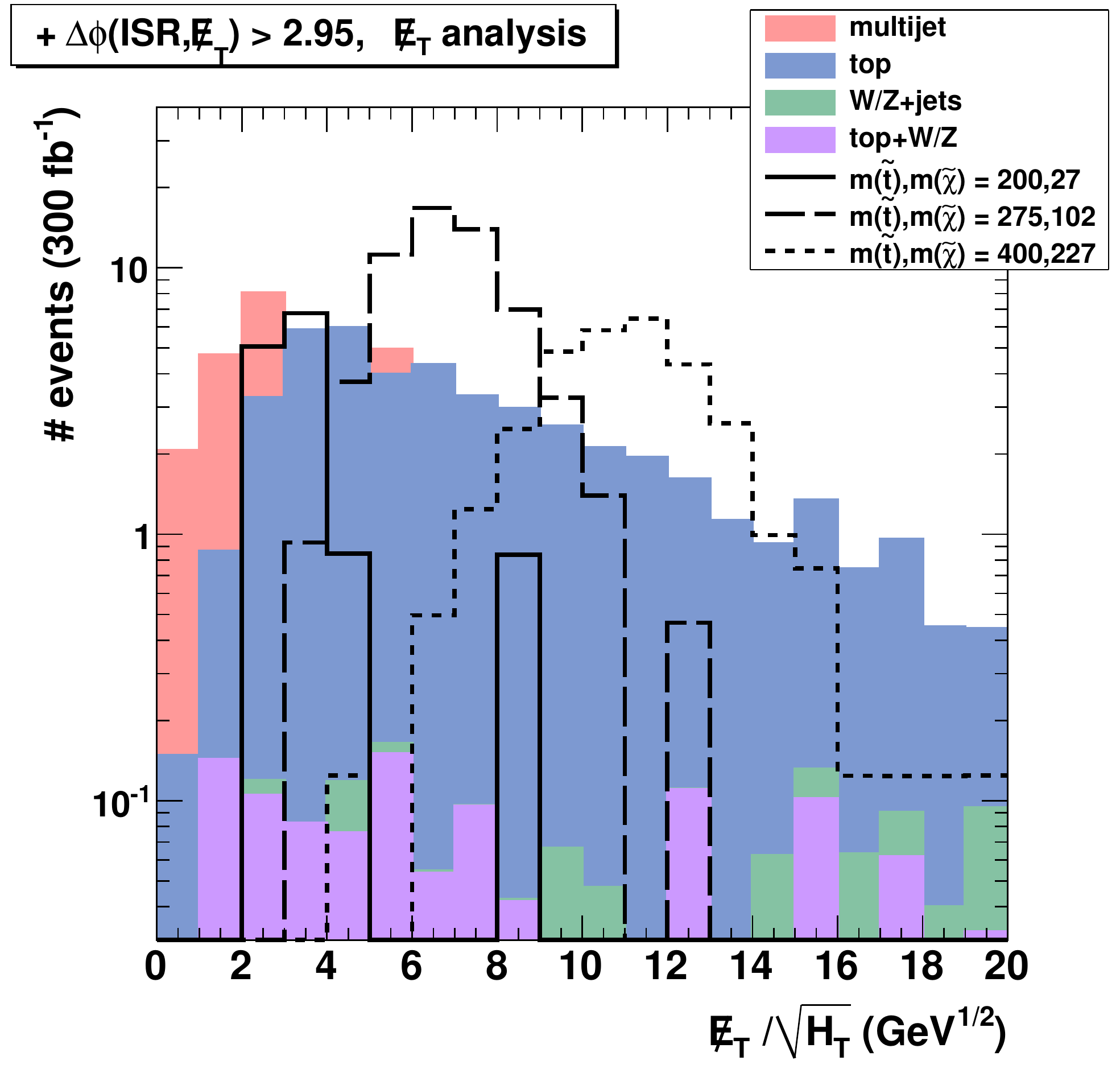}
\caption{Distribution of $\met$-significance for stacked backgrounds and some example signal points, for the $\mht$ analysis (left) and the truth-$\met$ analysis (right). All other cuts have been applied.}
\label{fig:metsig}
\end{center}
\end{figure*}

Table~\ref{tab:cuts} summarizes the complete set of cuts. Figs.~\ref{fig:mt1} and~\ref{fig:metsig} show distributions of several of the discriminating variables for backgrounds and some example signal points, illustrating the cumulative purification of the signal. Note that, to maintain efficient Monte Carlo generation, a cut of $\mht/\sqrt{H_T} > 3$~GeV$^{1/2}$ has been applied to define a baseline reconstructed sample (regardless of the final $\met$ definition used).

\begin{figure*}[tp!]
\begin{center}
\includegraphics[width=0.55\textwidth]{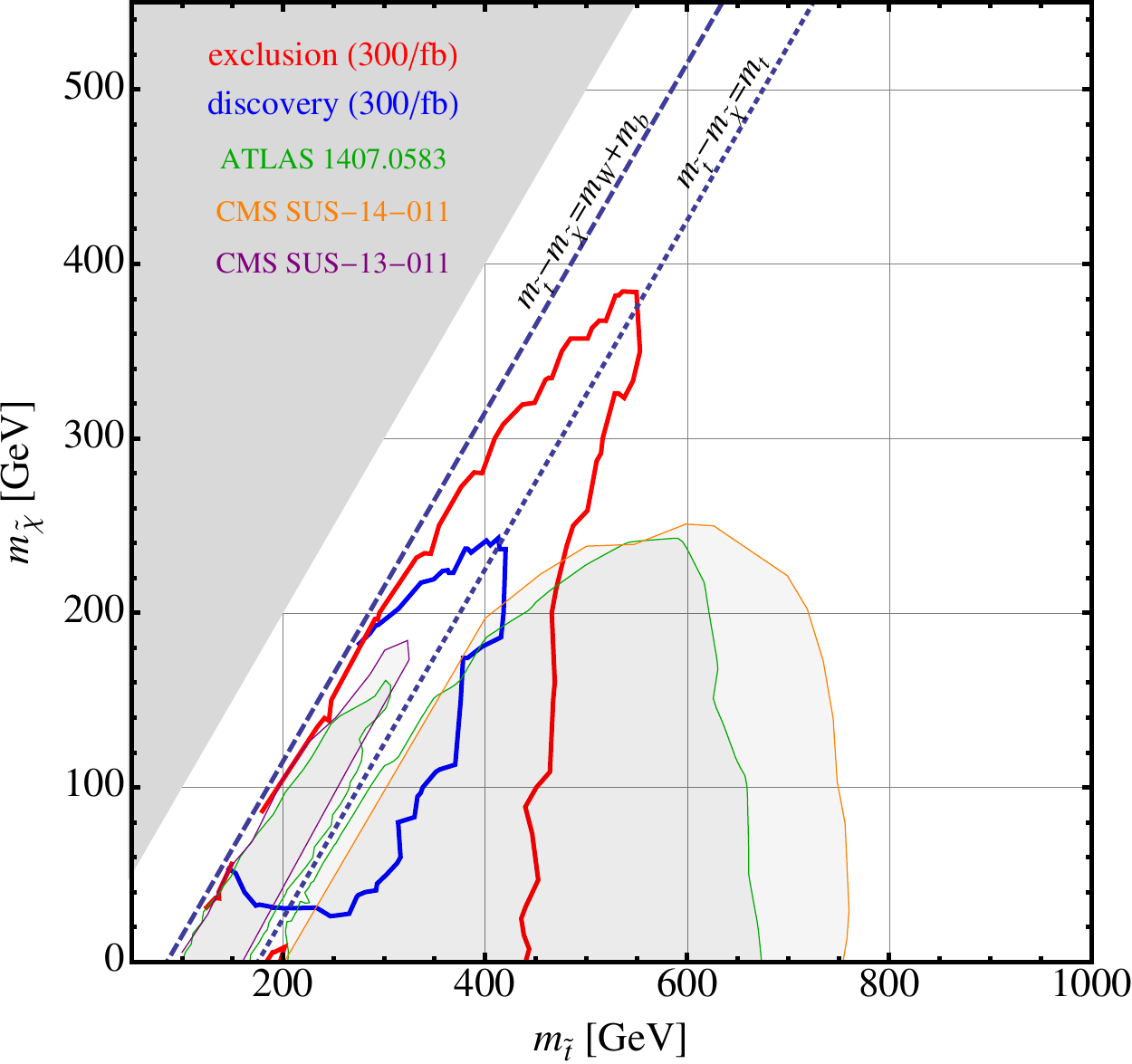}
\caption{Existing Run~1 limits from ATLAS and CMS, and projected 300~fb$^{-1}$ discovery and exclusion sensitivities for our $\mht$ analysis. The truth-$\met$ analysis (not shown) yields very similar exclusion contours, but somewhat stronger discovery contours at lower masses. Note that our simulation grid does not extend all the way down to the $W$ compression line $m_{\tilde t} \simeq m_{\tilde\chi} + m_b + m_W$ nor below, where the decay kinematics transitions to four-body. (We also do not indicate existing exclusions in that region. For the stealth exclusions, see Fig.~\ref{fig:stealth}.)}
\label{fig:sensitivity}
\end{center}
\end{figure*}

\begin{figure*}[tp!]
\begin{center}
\includegraphics[width=0.6\textwidth]{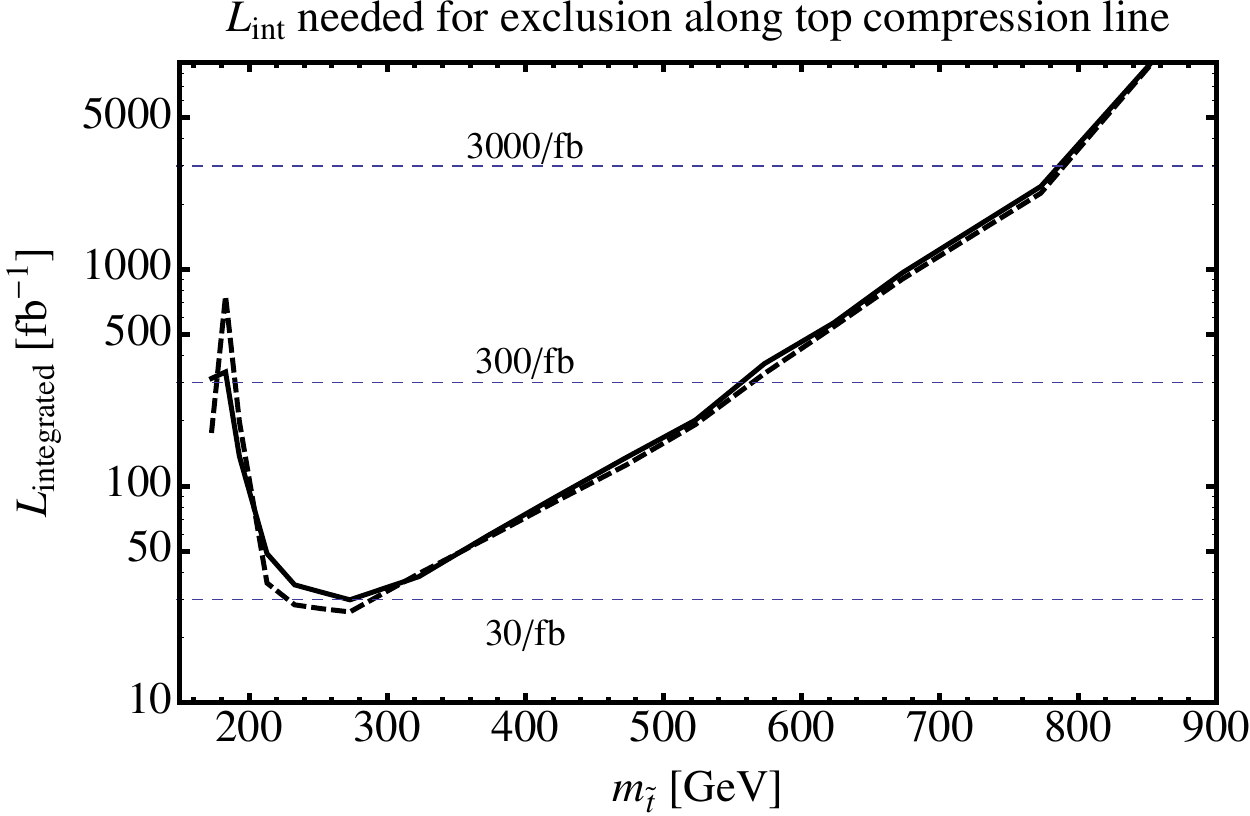}
\caption{Luminosity required for exclusion sensitivity along the top compression line for both our truth-$\met$ (dashed) and $\mht$ (solid) analyses, assuming 13~TeV and Run~2 \&~3 pileup and detector conditions. (Projections beyond 300~fb$^{-1}$ are naive extrapolations, not using HL-LHC conditions.)}
\label{fig:luminosity}
\end{center}
\end{figure*}

\begin{figure*}[tp!]
\begin{center}
\includegraphics[width=0.6\textwidth]{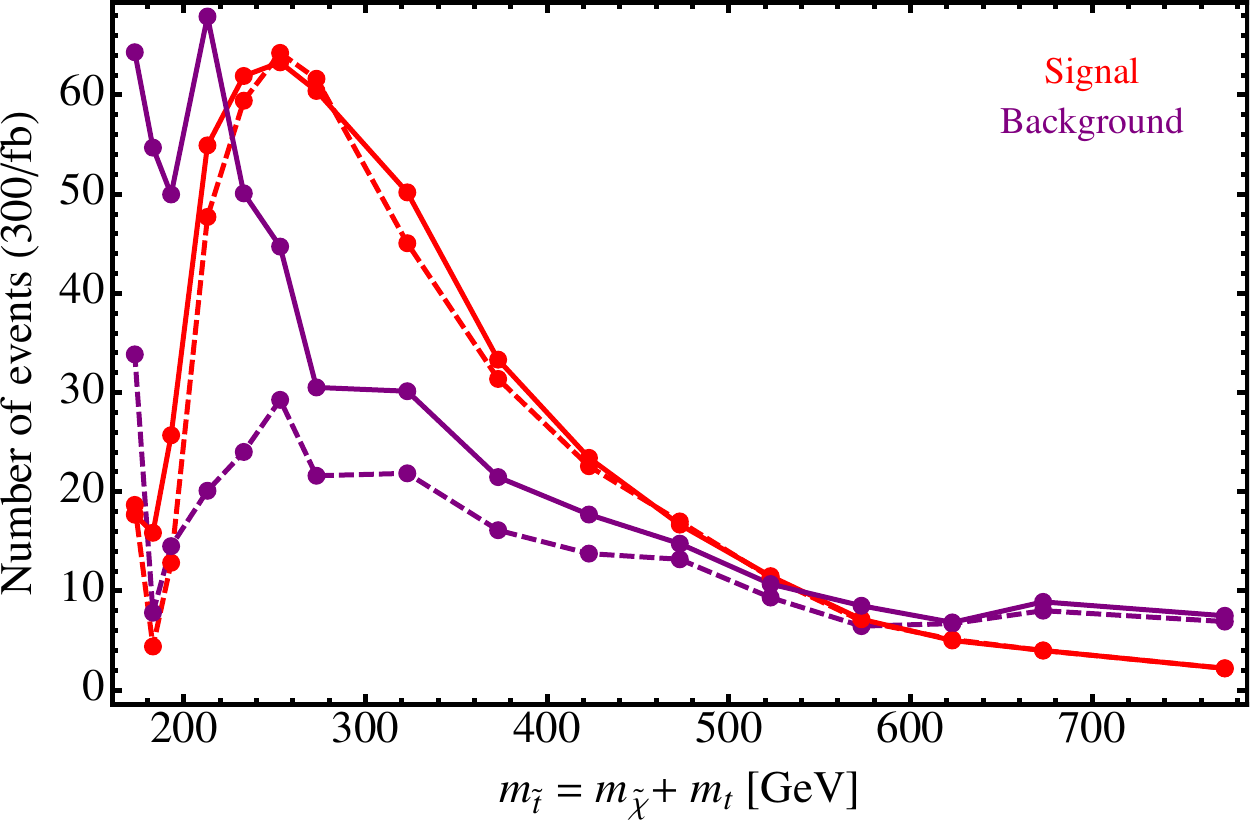}
\caption{Signal (red) and background (purple) counts along the top compression line for 300~fb$^{-1}$, for both our truth-$\met$ (dashed) and $\mht$ (solid) analyses.}
\label{fig:SandB}
\end{center}
\end{figure*}

\begin{figure*}[tp!]
\begin{center}
\includegraphics[width=0.6\textwidth]{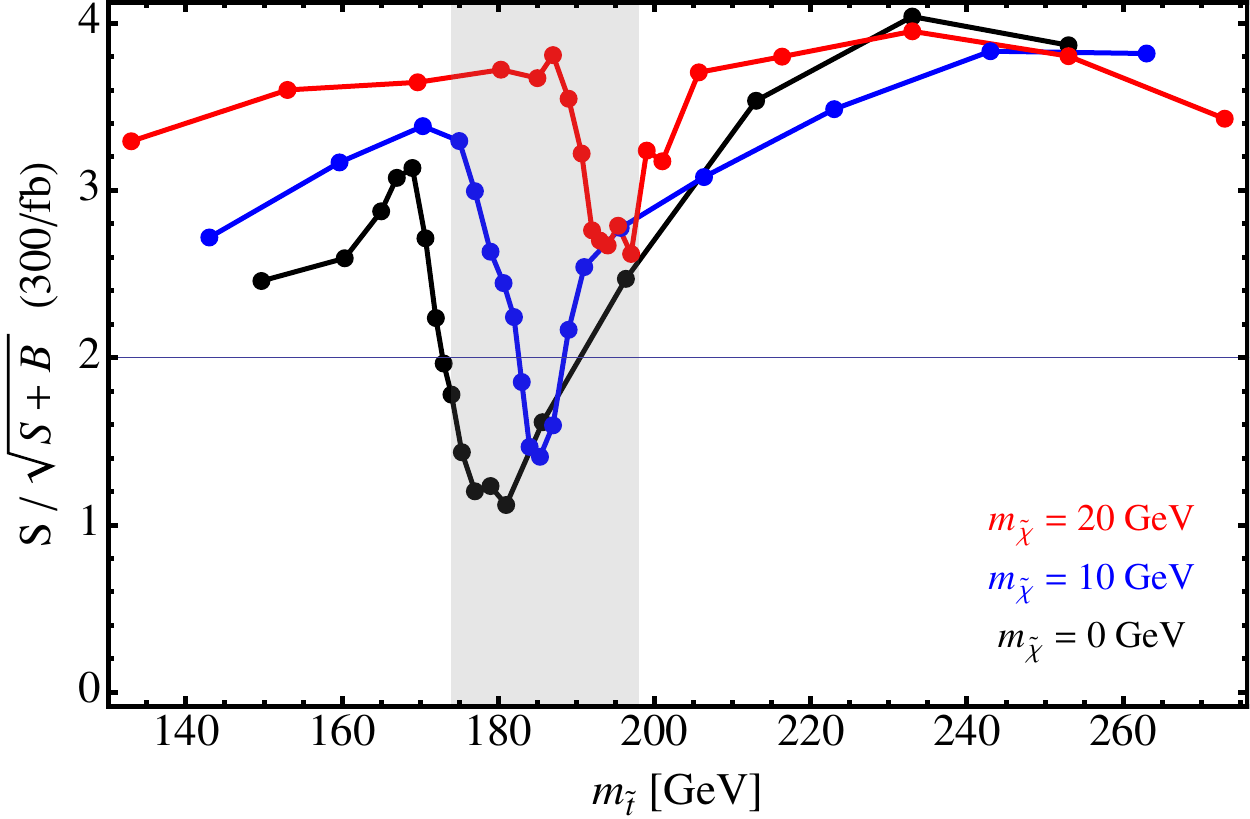}
\caption{Projected 300~fb$^{-1}$ exclusion sensitivity around the stealth point, scanning $m_{\tilde t}$ for several fixed $m_{\tilde\chi} \ll m_t$, for our $\mht$ analysis. The gray shaded region indicates the excluded stop masses from the dedicated ATLAS search for stealth stops \cite{Aad:2014mfk}. The truth-$\met$ analysis (not shown) yields very similar significances in the dip near the top compression line, but up to $O$(1) higher significances away from it.}
\label{fig:stealth}
\end{center}
\end{figure*}

The analysis thus defined, we scan through the model space of the stop-neutralino mass plane, with finer-grained steps near the top compression line (of order the top quark width). The final $\met/\sqrt{H_T}$ window is optimized per sample to maximize the naive statistical significance $S/\sqrt{S+B}$. We define exclusion threshold as $S/\sqrt{S+B} = 2$, and discovery threshold as $S/\sqrt{B} = 5$. Fig.~\ref{fig:sensitivity} shows our final exclusion and discovery contours for 300~fb$^{-1}$, indicating a near complete closure of the current compression line gap. Fig.~\ref{fig:luminosity} shows the luminosity required to achieve exclusion-level sensitivity along the compression line. While our simulations are done under Run~2 \&~3 conditions, we have also naively extrapolated as far as the HL-LHC luminosity of 3~ab$^{-1}$. We include as well in Fig.~\ref{fig:SandB} a scan of the signal and background rates at 300~fb$^{-1}$ along the top compression line. This indicates $S/B \sim 1$ over most of the range that we study, suggesting good resilience to systematic errors, which we have not attempted to estimate. Finally, in Fig.~\ref{fig:stealth} we provide a closer view of the exclusion sensitivity near the stealth point, via a series of scans over $m_{\tilde t}$ at fixed neutralino masses.

\section{Discussion and Outlook}
\label{sec:conclusions}

Fig.~\ref{fig:sensitivity} suggests that our proposed search strategy can access stops along the top compression line beyond 400~GeV at discovery-level significance, and perhaps up to 550~GeV at exclusion-level significance, over the current phase of LHC running. These numbers already start to approach what was done for non-compressed stops at Run~1. However, unlike those searches, for us the sensitivity is maximized {\it on} the top compression line. This complementarity is made possible by focusing on the unique kinematic configurations that start to open up at Run~2. It is rather remarkable that the sensitivity gap at the top compression line, which has become a modern benchmark of difficulty in new physics searches, can be shrunk so dramatically by simply going to a higher collider energy. Fig~\ref{fig:luminosity} indicates that the gap will start to close already with a data set comparable in size to Run~1, which should be achievable before the end of 2016.

On the low side, our search very closely surrounds the stealth point $(m_{\tilde t},m_{\tilde\chi}) = (m_t,0)$, as indicated in detail in Fig.~\ref{fig:stealth}. In fact, we have found that the exclusion-level contour there depends only moderately on whether we use truth $\met$ or $\mht$, though Fig.~\ref{fig:SandB} illustrates that this choice does strongly effect the $S/B$ there. We emphasize the caveat that we have not folded in systematic errors. Ultimately, the major question is how well the multijet background can be controlled and modeled. Given this uncertainty, it is difficult for us to make very concrete statements near the stealth point. But following the discussion in the introduction, it seems highly likely that multiple search strategies will come into play. Even the present state-of-the art searches based on $t\bar t$ cross section and spin correlation measurements~\cite{Aad:2014mfk,Aad:2014kva} already overlap with our projections, completing the coverage at exclusion-level.

Our search is also very effective at covering large portions of the three-body region. While our simulated model points do not extend below the $W$ compression line at $m_{\tilde t} \simeq m_{\tilde\chi} + m_b + m_W$, and into the four-body region, it seems quite likely that we even continue to have some coverage there. This leaves open the possibility of linking up with monojet and other searches in that region. (See as well~\cite{Kilic:2012kw} for a recast of a soft dilepton search at 7~TeV that already makes some surprising inroads there.) An approach that requires fewer jets and looser hadronic top reconstructions would also likely be fruitful, a possibility that we save for future work.

More generally, we have only very coarsely optimized our analysis, first by fixing most of our selection criteria by-eye on a small subset of model points, and then by selectively scanning over only our final $\met/\sqrt{H_T}$ window. With the principle proven, a more carefully optimized suite of cuts would certainly achieve better results, especially for the stealthier model points. Breaking the search into more analysis regions, e.g. binned over $p_T$(ISR-jet) (or fit over multiple variables), could also be beneficial. 

An obvious further extension of the analysis includes HL-LHC, with up to 3~ab$^{-1}$ of luminosity. The very high pileup would likely be a major concern there, as the rate of fake jets rises significantly, and the resolution on $\met$ further degrades. Certainly, pushing further into the stealth region will be difficult, although the much higher event rates may allow for more highly-crafted cuts. On the high-mass side, if we naively extrapolate up our 300~fb$^{-1}$ analysis as per Fig.~\ref{fig:luminosity}, we find discovery (exclusion) reach extending to about 800~GeV. Along similar lines, projections for a 100~TeV proton collider are also interesting to pursue. However, as we ultimately scan up to $m_{\tilde t} \gg m_t$, we effectively return to the fully compressed situation $m_{\tilde\chi} \simeq m_{\tilde t}$. All of the compression lines may then practically blur together using more standard ``monojet''+$\met$ style searches, perhaps supplemented by the additional ``soft'' activity from the $t^{(*)}$ decays. Such an analysis has been carried out in~\cite{Cohen:2014hxa}, finding sensitivity to compressed stops up to multiple TeV using the dilepton channel.\footnote{If we naively scale the energies and cross sections from the existing monojet+$\met$ searches for fully compressed stops~\cite{Aad:2014nra,CMSstopToCharm} from an 8~TeV machine to a 100~TeV machine (without running the PDFs), we would expect an exclusion of $(260~{\rm GeV}) \times (100/8) \simeq 3$~TeV after accumulating a luminosity of $20~{\rm fb}^{-1} \times (100/8)^2 \simeq 3$~ab$^{-1}$. Suggestively, this coarse estimate is very close to that of~\cite{Cohen:2014hxa} on the top compression line.} Finally, all of our results readily generalize to those classes of fermionic top-partner models that exhibit either a conserved or approximately-conserved parity, and contain a neutral ``LSP'' boson which plays a role kinematically identical to $\tilde\chi^0$~\cite{Cheng:2005as}. The only major difference relative to stops, from the perspective of our analysis, is their approximately six times larger cross section at a given mass, yielding commensurately stronger sensitivity.

In conclusion, natural supersymmetry poses some interesting phenomenological challenges, as evidenced by the enduring gaps in coverage of one its simplest incarnations: an NLSP stop and LSP neutralino. While limits continue to push upward in mass in the favorable parameter regions that readily provide lots of $\met$, we have seen here that an appropriately constructed analysis at the upgraded LHC, along the lines suggested in~\cite{Hagiwara:2013tva}, can qualitatively extend sensitivity to this model into the more difficult compressed regions at lower masses. Combined, these approaches will leave very little ``natural'' parameter space unexplored. With its next major phase in progress, the LHC appears poised to provide us with a much more comprehensive perspective on the possible role of supersymmetric top quarks in Nature.

\vspace{0.8cm}

{\bf Note added:} While this paper was nearing completion,~\cite{An:2015uwa} appeared, which has significant overlap with our results. Their proposed $R_M$ variable (a very close variant of what was originally proposed in~\cite{Hagiwara:2013tva}) is highly correlated with the $\met/\sqrt{H_T}$ variable that we use here, and in general with any variable proportional to $\met$ in the presence of a hard ISR-jet $p_T$ cut. There are a number of other differences in our analysis strategy, which lead to a higher $S/B$ with comparable formal statistical significance, and somewhat different sensitivity contours. We also pay additional attention to the approach to the stealth region and the possible role of $\met$ resolution. However, we do not make a dedicated study around the $W$ compression line.


\acknowledgments{We thank Ayres Freitas, Matthew Buckley, Kaoru Hagiwara, Eva Halkiadakis, George Redlinger, and Scott Thomas for discussions, and Olivier Mattelaer for help debugging MadGraph matching at high $k_T$. SM was supported by NSF grant No.\ PHY-1404056.  MP was supported by the LHC-TI grant NSF No.\ PHY-0969510 and by the VPUE at Stanford University.  DS was supported by DoE grant No.\ DE-SC0003883.  BT was supported by DoE grant No.\ DE-FG02-95ER40896 and by PITT PACC.}


\appendix

\section{Event Generation}
\label{sec:generation}

Our event generation is performed using {\tt MadGraph5\_aMC@NLO}~\cite{Alwall:2014hca} at 13~TeV and showered with {\tt PYTHIA~6}~\cite{Sjostrand:2006za}, using leading-order matrix elements (without K-factors). We set the top quark mass to 173~GeV, and width to 1.5~GeV.

For our signal samples, we choose mostly-right-handed stop and mostly-Bino neutralino (spin effects on our all-hadronic analysis are expected to be modest). Most samples are generated as $\tilde t \tilde t^* j$, with only a parton-level cut of 400~GeV on the accompanying jet. Both stops are decayed using three-body phase space $\tilde t \to Wb\tilde\chi^0$, regardless of mass point, which is crucial for modeling the kinematic transition at the top compression line. A complete decay chain is therefore, e.g., {\tt t1~$>$ W+ b n1, W+~$>$ j j}. The stop width for each model point is computed separately using $1 \to 3$ parton-level decay simulations. A subset of models along the compression line have been simulated over their full production phase space, using $k_T$-MLM matching with a threshold of 100~GeV. Perhaps unsurprisingly, the events passing our final selections are highly dominated by the $1j$ subsample, and are in close agreement with our simple unmatched simulations. Similarly, we find very low relative pass rates for decay modes other than all-hadronic.

The backgrounds are generated as follows.\footnote{We do not generate single-top nor diboson backgrounds, which, given our reconstruction criteria, we expect to be subdominant to $t\bar t$ and $W/Z$+jets, respectively.  Cf.\ the background composition in ATLAS's all-hadronic stop search~\cite{Aad:2014bva}.} Our $t\bar t$ sample is matched up to one (two) jets for all-hadronic (partially leptonic or $\tau$) decays, again using a 100~GeV threshold. We also generate $t\bar t W$ and $t\bar t Z$ matched up to one jet. For $W/Z$+jets and multijet backgrounds, we concentrate on production with at least two heavy quarks (bottom or charm) in the hard event. Because of the difficulties of computing very high-multiplicity matrix elements, we mainly use the parton shower to generate extra partons, and do not employ any matching. The $W/Z$+jets sample specifically starts with $W/Z$ (decaying to $l\nu$, $\tau\nu$, or $\nu\nu$) plus three hard partons, while the multijet sample starts with four hard partons. We have also cross-checked the multijets against {\tt AlpGen}~\cite{Mangano:2002ea} samples, generated with identical criteria. For each sample we impose cuts at the parton level that treat the $b$ and $j$ partons democratically, requiring $p_{T}(j) > 15$, $\Delta R(j,j) > 0.4$ (where $j$ here includes $b$) as well as a $p_T$ cut on the hardest jet of 350~GeV.


\bibliography{lit}
\bibliographystyle{apsper}

\end{document}